\documentclass[a4paper,10pt]{article}
\usepackage{amsmath, amssymb,natbib}
\usepackage[utf8]{inputenc}
\usepackage{graphicx,pstricks,epstopdf}
\usepackage{graphics}
\usepackage{subfigure}
\usepackage{epsfig}
\usepackage{subfigure}
\usepackage{txfonts}
\usepackage{palatino}
\usepackage{rotating}
\usepackage{bm}
\usepackage[margin=1.15in]{geometry}
\usepackage{ragged2e}

\title{A Scalable Empirical Bayes Approach to Variable Selection}

\author{Haim Y. Bar\thanks{Haim Bar is an Assistant Professor, Department of
Statistics, University of Connecticut (e-mail:
haim.bar@uconn.edu).}\and
James G. Booth\thanks{James Booth is Professor, Department of Biological
Statistics and Computational Biology, Cornell University, Ithaca NY, 14853
(email: jb383@cornell.edu).  Professor Booth's research was partially supported
by an NSF grant, NSF-DMS 1208488.} \and
Martin T. Wells\thanks{Martin T. Wells is Professor, Department of Biological
Statistics and Computational Biology
and the Department of Statistical Science, Cornell University, Ithaca NY, 14853
(email: mtw1@cornell.edu). Professor Wells' research was partially supported by
NSF-DMS 1208488 and NIH grant U19 AI111143.}}

\begin{document}

\maketitle

\begin{abstract}
We develop a model-based empirical Bayes approach to variable selection problems
in which the number of predictors is very large, possibly much larger than the number
of responses (the so-called “large p, small n” problem). We consider the multiple linear 
regression setting, where the
response is assumed to be a continuous variable and it is a linear function of
the predictors plus error. The explanatory variables in the linear model can have a positive
effect on the response, a negative effect, or no effect. We model the effects
of the linear predictors as a three-component mixture
in which a key assumption is
that only a small (unknown) fraction of the candidate predictors have a non-zero effect on
the response variable. By treating the coefficients as random effects we develop
an approach that is computationally
efficient because the number of parameters that have to be estimated is small, and
remains constant regardless of the number of explanatory variables. 
The model parameters are estimated using the EM algorithm which
is scalable and leads to significantly faster convergence, compared with simulation-based methods.
\end{abstract}

\section{Introduction}
This manuscript focuses on variable selection in normal linear regression models
when there are a large number of candidate explanatory variables, most of which
have little or no effect on the dependent variable. We propose an empirical
Bayes, model-based approach to variable selection which we implement via a fast
EM algorithm.

Traditional regression problems typically involve a small number of explanatory
variables and an analyst can make educated decisions as to which ones should be
included in the regression model, and which should not. However, the new age of
high speed computing  and technological advances in
genetics and molecular biology, for example, have dramatically changed the modeling and computation paradigms. It is common
practice to use linear regression models to estimate the effects of hundreds or
even thousands of predictors on a given response.  These modern applications
present major challenges. First, there is the so-called `large $p$, small $n$'
problem, since the number of predictors, $p$, e.g. gene expressions from a microarray or RNASeq
experiment, often greatly exceeds the sample size, $n$.
Methods controlling the experiment-wise false discovery rate in one predictor at
a time analyses often result in few or no discoveries.
Second, the model space is huge. For example, for a modest QTL study with 1000
markers, there are $2^{1000}$ possible models.  This renders exhaustive
search-based algorithms impractical.

Automated methods for variable selection in normal linear regression models have
long been studied in the literature \citep{Hocking:1976,Breiman:1995,casella:2006,
george:1993}.  Virtually every statistical package contains an implementation of standard
stepwise methods.  These stepwise  methods typically add or remove one variable from the model in each
iteration, based on sequential F-tests and a threshold, or based on the change in other
goodness-of-fit type measures, including adjusted R-square, AIC , BIC, or $C_p$.
Other approaches use the false discovery rate (FDR) procedure \citep{benjamini:hochberg}. See for example, 
\cite{benjamini2009}.

A number of model selection procedures are based criteria that incorporate a penalty based on model size in
order to discourage complexity \citep{boisbunon:2014}.  
Akaike information
criterion (AIC) \citep{Aikake:1974} and Bayesian information criterion (BIC)
\citep{Schwarz:1978} belong to this family.
The Mallows' $C_p$ \citep{mallows:1973} statistic is
also similar in purpose.

Other approaches include variations of penalized likelihood approaches
in which the
coefficient estimation and variable selection are done simultaneously.  The $\ell_0$ norm, which counts the number of non-zero elements in a vector,
underlies the mathematical foundations of all sparsity related problems.  The $\ell_0$ norm is not convex and thus the
convergence to the global optimum is hard to ensure. Minimizing the $\ell_0$ norm penalization is NP-hard and thus is
often relaxed to the $\ell_1$ norm, the sum of the absolute values of the vector.   In the context of compressed sensing
\cite{candes2005} showed that when a restricted isometry property of the design matrix holds, the $\ell_0$ norm
minimization can be equivalently replaced by its convex relaxation $\ell_1$ norm.  The LASSO \citep{Tibshirani:1996}
minimizes the residuals sum of squares subject to an $\ell_1$ constraint. This constraint ensures that the number
of non-zero parameter estimates is controlled and adapts to sparsity. Other methods that are based on a minimizing
a loss function, subject to a constraint on the complexity of the model include SCAD - the smoothly clipped
absolute deviation \citep{FanLi:2001}, the adaptive LASSO \citep{Zou:2006}, 
LARS - Least Angle Regression \citep{Efron:2004,Hesterberg:2010}.
and more recently, 
\cite{Malgorzata2015} proposed the Sorted $\ell_1$ Penalized Estimator (SLOPE) for the vector of regression coefficients in the linear regression
setting where $p$ may be greater than $n$.

A major shortcoming of penalized likelihood approaches, in particular the LASSO, is the need to specify a tuning parameter that
is properly adjusted to all aspects of the proposed model and therefore difficult to implement in practice \citep{LedererM15}.
Using cross-validation to select the tuning parameter is not a satisfactory approach since cross-validation is
computationally inefficient and provides unsatisfactory variable selection performance \citep{hebiri:2013}.  Furthermore,
correlations in the design matrix also influence Lasso prediction.  \cite{hebiri:2013} find that the larger the correlations, the smaller the optimal
tuning parameter. Consequently the model selection is less penalized giving rise to a less parsimonious model.

Other variations on the LASSO approach include \cite{buhlmann2014} and \cite{LedererM15}. Specifically, \cite{LedererM15}
introduce an alternative to LASSO with an inherent calibration to all aspects of the model. This adaptation leads to an
estimator that does not require any calibration of tuning parameters and deals well with correlations in the design matrix.
These two papers demonstrate the performance of several methods when applied to
 a high-dimensional data set which involves 4,088 genes and only
71 subjects.  We apply our method to this data set and compare our results with the ones
obtained by using the \texttt{hdi} package \citep{hdi:2014} and B-TREX \citep{LedererM15}.

Our method is more related to Bayesian approaches which include
\cite{casella:2006} and \cite{george:1993}, and the spike-and-slab method \citep{Ishwaran:2005}.  Spike-and-slab priors are linked to the $\ell_0$ penalty in that a spike at zero is connected to the number of zero components of the regression parameter.  More recently, \cite{Bondell:2012}, among others, considered model selection consistency in Bayesian variable selection methods.
Our motivation and modeling approach is similar to \cite{zhang:2005} and \cite{Guan:2012} who
also use a Bayesian approach and whose work is motivated by QTL and gene-wise association studies (GWAS).
Our model-based approach allows for a fully-Bayesian implementation, but we use an empirical
Bayes approach because the running time of an MCMC sampler is too long for many data sets
in modern applications.

We model the effects of the linear predictors as a three-component mixture
where only a small (unknown) number of predictors have a non-zero effect on the outcome.
Implementation of the empirical Bayes approach is not trivial for two reasons.
First, the likelihood depends on a large number of latent indicator variables which
determine which variables have a non-zero effect on the response, and hence are included in the 
linear model. We treat these variables as missing data
and use the EM algorithm \citep{Dempster1977}. However, the E-step is analytically intractable
because the complete data log-likelihood is a non-linear function of the latent variables.
Furthermore, the complete data likelihood function involves the inverse of a matrix, which in this
large $p$ small $n$ setting is approximately singular.
We address these problems in Section \ref{sec:vs:est}. The first problem is solved by deriving
an approximate EM algorithm using Bayes' rule to estimate the posterior probabilities of the
latent variable. The second problem is solved by using the Woodbury identity \citep{Golub:1996} which
  results in a
non-singular, low rank variance-covariance matrix. The approximate EM algorithm,
combined with the Woodbury identity  results in much improved performance.
See \cite{muller2013}
for a discussion and comparison of other model selection methods based on AIC, BIC,
shrinkage methods based on LASSO penalized loss functions and Bayesian
techniques in the context of linear mixed models.

This article is organized as follows.  We introduce the model-based approach in
Section \ref{sec:vs:model}.  In Section \ref{sec:vs:est} we derive the complete
data likelihood function, the EM algorithm procedure, and the selection procedure.
In Section \ref{sec:implementation} we discuss some important
computational considerations.
In Section \ref{sec:sim} we show results from a simulation study, and in Section
\ref{sec:vs:cases} we illustrate our method using two well-known data sets and compare
our results with others in the literature.
In Section \ref{sec:varsel:ext} we discuss
extensions to the model, and we conclude with  Section \ref{sec:varsel:conc}.

\section{A Statistical Model for Automatic Variable
Selection}\label{sec:vs:model}
We deal with a multiple linear regression setting
in which the number of predictors is
large and in some cases even much larger than the sample size.
To construct our model we begin with some notation and assumptions.
Denote the continuous responses by $y_i$, $i=1,\ldots,N$, and assume that the
mean response  is a linear function of several predictors. We allow for $J\ge 0$ predictors (e.g. sex, population, age) 
that are always included in the model and $K>0$ `putative' variables, of which
only a small subset should be included. We may not have
any prior knowledge as to which putative variables are associated with
the response. For example, we may be interested in the association between
gene expression levels that are available for thousands of genes, and a
quantitative trait such as body-mass index (BMI).

Denote the $j$th `locked in' variable by $\mathbf{x}_{j}$, and let the mean effect of the
$j$-th covariate be $\beta_{j}$.
Denote the $k$th putative variable by $\mathbf{z}_{k}$.
We assume that the response $y_i$ can be modeled using an additive combination
of the predictors as follows:

\begin{eqnarray}\label{model_1}
y_{i} &=&\sum_{j=1}^{J}x_{ij}\beta _{j}+\sum_{k=1}^{K}z_{ik}\gamma _{k}u
_{k}+\varepsilon _{i}
\end{eqnarray}
where
\begin{eqnarray*}
u_{k} &\stackrel{iid}{\sim} &N\left( \mu ,\sigma ^{2}\right)  \\
\gamma _{k} &\stackrel{iid}{\sim} &multinomial\left(0,1,-1; p_0,p_1,p_2\right)
\\
\varepsilon _{i} &\stackrel{iid}{\sim} &N\left( 0,\sigma _{e}^{2}\right)\,.
\end{eqnarray*}
The multinomial random variables $\{\gamma_k\}$ take the value 1 or -1 if and only
if the $k^{th}$ putative variable is included in the model. Its sign
indicates whether the mean effect of the $k^{th}$ variable on the response is
positive or negative.
In this context, the problem of variable
selection can be seen as an
estimation procedure, where the main interest is identifying which of the latent variables
$\{\gamma_k\}$ are non-zero.

For the derivation of the parameter estimates it is convenient to express
the model in matrix notation.  Denote the $N\times
K$ matrix $\left( z_{ik}\right) $ by $\mathbf{Z},$
and write $\bm{\Gamma} \equiv diag\left( \gamma _{1},\gamma
_{2},\ldots,\gamma _{K}\right)$ and $\bm{\mu}=\mu\mathbf{1}_K$.  Let
$\mathbf{z}_k$ denote the $k^{th}$ column of $\mathbf{Z}$.  Also, denote the
$N\times J$ matrix
$\mathbf{X} =(x_{ij}),$ and the $J\times 1$ vector of fixed effects
$\bm{\beta} =(\beta_1,...,\beta_J)'.$
Then the model can be rewritten as
\begin{eqnarray}\label{model:varsel}
\mathbf{y}
&=&\mathbf{X}\bm{\beta}+\mathbf{Z}\bm{\Gamma}\mathbf{u}+\bm{\varepsilon}\\
\bm{\varepsilon } &\sim &N\left( \mathbf{0}_{N},\sigma _{e}^{2}%
\mathbf{I}_{N}\right) \\
\mathbf{Z}\bm{\Gamma}\mathbf{u}\,|\,\bm{\Gamma} &\sim &N\left(
\mathbf{Z}\bm{\Gamma}\bm{\mu},\sigma ^{2}\mathbf{Z}\bm{\Gamma}^2%
\mathbf{Z}'\right)\,,
\end{eqnarray}
which is similar to the usual mixed-model representation, but with two notable
differences.  First, the model includes the diagonal matrix, $\bm{\Gamma}$,
which is used to select the columns from $\mathbf{Z}.$  Second, the mean of the
random effect terms is not zero.  Note that in the usual mixed model context,
the mean of the random effect is not identifiable separately from the overall
mean, and therefore it is assumed to be 0. However, in mixture models (e.g.
\citealt{BaBoScWe2010}) this is not the case, and in fact, not only are the two
means identifiable, letting $\mu$ be non-zero allows us to separate the
significant covariates into two groups (positive and negative mean effect)
resulting in an increase in power (compared with the two-group mixture model).

\section{Estimation}\label{sec:vs:est}
\subsection{The Complete Data Likelihood}

We use an empirical Bayes approach in which the parameters
$\theta=\{\bm{\beta}, \mu, \sigma_e^2, \sigma^2, \bm{p}
\}$
are estimated via a modified
EM algorithm in which we treat the indicators
  $\{\gamma_k\}$ as missing values. Upon convergence we select a column $\mathbf{z}_k$ to be
included in the model if the estimated posterior probability of its latent
indicator, $\gamma_k$, is greater than a predefined threshold.
The complete data likelihood, $f_C(\mathbf{y},\bm{\Gamma})$, is obtained by
integrating out the random effects, $\{u_k\}$. Then the $Q$-function for the EM
algorithm is given by $Q(\theta|\theta^{(t)}) = E_{\theta^{(t)}}\{\log
f_C(\mathbf{y},\bm{\Gamma})|\mathbf{y}\}.$

Define
\begin{eqnarray*}
c_0 &=& \sum_{k=1}^{K}I\left[ \gamma _{k}=0\right]\\
c_1 &=& \sum_{k=1}^{K}I\left[ \gamma _{k}=1\right]\\
c_2 &=& \sum_{k=1}^{K}I\left[ \gamma _{k}=-1\right]
\end{eqnarray*}
where
$I\left[ \cdot\right]$ is the indicator function.  
Then the (complete data) log-likelihood
function is given by
\begin{eqnarray}\label{log.likelihood}
\ell=\log f_C(\mathbf{y},\Gamma) &=&c_0\log(p_{0})+c_1\log(p_{1})+
c_2\log(p_{2})-\frac{N}{2}\log\left( 2\pi
\right)\nonumber\\
&&-\frac{1}{2}\log\left\vert
\sigma_{e}^{2}\mathbf{I}_{N}+\sigma^2\mathbf{Z}\bm{\Gamma}^2 \mathbf{Z}'
\right\vert \nonumber\\
&&-\frac{1}{2}
\left( \mathbf{y}-\mathbf{X}\bm{\beta}-\mathbf{V} \bm{\mu }\right)'
( \sigma_{e}^{2}\mathbf{I}_{N}+\sigma^2\mathbf{Z}\Gamma^2 \mathbf{Z}')^{-1}
\left( \mathbf{y}-\mathbf{X}\bm{\beta }-\mathbf{V}\bm{\mu }\right)\label{eq:log.lik}\,.
\end{eqnarray}
The likelihood function is simply the probability distribution
function of a multivariate normal random variable with mean $\mathbf{X}\bm{\beta
}+\mathbf{V}\bm{\mu }$ and variance-covariance matrix
$\bm{\Sigma}=\sigma_{e}^{2}\mathbf{I}_{N}+\sigma^2\mathbf{Z}\Gamma^2
\mathbf{Z}'$, multiplied by the prior probability of the latent variables.

\subsection{The EM Algorithm}
We fit the model using the Expectation Maximization (EM) algorithm \citep{Dempster1977}
except that in the M-step we plug in the
current estimates of the posterior expected
values of the latent variables $\gamma_k$ and maximize with respect to $\mu, \bm{\beta}, \sigma^2,$
and $\sigma_e^2$. In the E-step we use the current  estimates
of $\mu, \bm{\beta}, \sigma^2,$ and $\sigma_e^2$ and compute the expected values
of $\gamma_k$. We repeat the process until convergence is reached. For example,
the convergence criterion can be defined in terms of change in the log-likelihood
function.

The M-step:
denote $\mathbf{W}=\left[\mathbf{X},\mathbf{Z}\bm{\Gamma}\mathbf{1}_K\right].$
The mean of the multivariate normal distribution in the complete data likelihood
is $\mathbf{W}\bm{\tilde{\beta}}$ where $\bm{\tilde{\beta}}=(\bm{\beta}',\mu)'.$
 Then the MLE for $\bm{\tilde{\beta}}$ is given by
\begin{eqnarray}
(\hat{\bm{\beta}}',\hat{\mu})'&=&
(\mathbf{W}'\bm{\Sigma}^{-1}\mathbf{W})^{-1}
\mathbf{W}' \bm{\Sigma}^{-1}\mathbf{y}\,.\label{mle.eq.beta}
\end{eqnarray}%

To estimate the variance parameters we use the following equations (see Section
8.3.b in \citealt{McCulloch:1992}).
Using the values from the $t^{th}$ iteration of the EM algorithm, define
\begin{eqnarray*}
\tau_e&=&trace[\sigma_e^2\mathbf{I}_N-\sigma_e^4\bm{\Sigma}^{-1}]+
\sigma_e^4(\mathbf{y}-\mathbf{W}\bm{\tilde{\beta}})'\bm{\Sigma}^{-2}
(\mathbf{y}-\mathbf{W}\bm{\tilde{\beta}})\\
\tau_r&=&trace[\sigma^2\mathbf{I}_K-\sigma^4\mathbf{V}'
\bm{\Sigma}^{-1}\mathbf{V}]+
\sigma^4(\mathbf{y}-\mathbf{W}\bm{\tilde{\beta}})' \bm{\Sigma}^{-1}
\mathbf{V}\mathbf{V}'
\bm{\Sigma})^{-1}(\mathbf{y}-\mathbf{W}\bm{\tilde{\beta}})
\,.
\end{eqnarray*}
Then the $t+1$ updates to the variance terms are
\begin{eqnarray}
\sigma_e^2&=&\frac{\tau_e}{N}\label{sigma_e}\\
\sigma^2&=&\frac{\tau_r}{rank(\mathbf{V})}\label{sigma}\,.
\end{eqnarray}
Maximizing with  respect to $p_{0},p_{1},p_2$ yields $p_{m}=c_m/K$.

The E-step: to update the latent variables $\gamma_k$ we use Bayes' rule to compute
the posterior probability that the $k$th putative variable is included in the model. For example:%
\begin{eqnarray}\label{postprob}
Pr\left(\gamma _{k}=-1\right)&=&\frac{p_{2}^{\left( t\right) }f\left(
\mathbf{y;}\gamma _{k}=-1,\gamma _{-k}=\gamma _{-k}^{\left( t\right)
}\right) }{\sum_{s\in\{-1,0,1\}}p_{i(s)}^{\left( t\right) }f\left( \mathbf{y;}\gamma
_{k}=s,\gamma _{-k}=\gamma _{-k}^{\left( t\right) }\right) }
\end{eqnarray}
where $f(\cdot)$ is the exponent of the log-likelihood in (\ref{log.likelihood}) given the current
parameter estimates, and $i(s)=0,1,2$ for $s=0,1,-1$, respectively. The notation
$\gamma _{-k}=\gamma _{-k}^{\left( t\right) }$ means that to update the $k^{th}$
component of the diagonal matrix $\bm{\Gamma}$ we hold all the other ones
constant at their value from the previous iteration.

\subsection{Variable Selection}\label{subsec:selection}
To control the number of putative variables that enter the regression model,
we propose two (related) methods to select a small number of variables that fit the
data well. One method is based on the posterior probabilities of the indicator variables
$\gamma_k$, and one is based on the likelihood-ratio between nested models.
The key idea is to obtain a sparse model by setting  $\gamma_k$ to 0 if
the posterior probability (or the likelihood) suggest that the $k$-th variable
has a weak association with the response.

Using the posterior probability method, we set $\gamma _{k}=0$ if
$Pr\left(\gamma _{k}=0\right)$ is greater than a
certain threshold.  Otherwise, if $Pr\left(\gamma _{k}=1\right) > Pr\left(\gamma
_{k}=-1\right)$ we set $\gamma _{k}=Pr\left(\gamma _{k}=1\right)$ and if
$Pr\left(\gamma _{k}=1\right) \le Pr\left(\gamma _{k}=-1\right)$ we set $\gamma
_{k}=-Pr\left(\gamma _{k}=-1\right)$.
In other words, we include the $k$-th covariate if and only if $Pr\left(\gamma
_{k}=0\right)$ is less than a certain threshold.  This will have significant
computational benefits when $N$ and $K$ are large and only a small number of
covariates have a significant effect on the response.  We refer to the variables
that are excluded from the model
(i.e. $\gamma_k=0$) as `\textit{null}'.

For the likelihood ratio method, denote the current value of
the log-likelihood by $\ell^t$ (given the current values of the parameter estimates).
Then, for each putative variable we compute the likelihood if $\gamma_k=0$,
1, or -1 and denote these values by $\ell_k(0)$, $\ell_k(1)$, and $\ell_k(-1)$.
Let $d_k(s)=\max \{\ell_k(s)-\ell^t ~|~ s\in\{0,1,-1\}\}$. If $d_k(s)~>~0$ then changing
$\gamma_k$ from its current value to $s_k^*=argmax\{\ell_k(s)-\ell^t\}$ will increase the
overall likelihood by $d_k(s_k^*)$.
If in the current iteration there are multiple variables for which $d_k(s_k^*)>0$,
modify $\gamma_k$ to equal $s_k^*$ for just one $k$, using one of two methods:
\begin{itemize}
 \item The greedy method: choose $k$ for which $d_k(s^*)$ is largest.
 \item The weighted probability method: choose $k$ with probability
 $$\frac{d_k(s_k^*)}{\sum_r d_r(s_r^*)}.$$
\end{itemize}
In some cases the increase in the likelihood may be quite small and not yield meaningful
improvement in the overall likelihood. A
modification of the algorithm is to only consider putative variables
for which $d_k(s_k^*)>\delta>0$. For
example, using $\delta=\log(2)$ means that only variables that yield a minimum two-fold increase in
the likelihood are considered for inclusion.

Thus, variable selection is achieved via the approximate
\textit{ maximum likelihood estimation} of the latent variables, and the estimation
of the posterior probabilities or the likelihood ratio criterion for inclusion in
the regression model requires a \textit{small, and fixed number of
parameters} which have to be estimated, regardless of the number of putative
variables.

\subsection{When N and K are Large -- the Modified EM Algorithm}
Application of the EM algorithm is not straightforward for two
reasons.  First, the complete data log-likelihood is a non-linear function of
the latent variables, making the E-step analytically
intractable.  We solve this
problem by updating the $\gamma_k$'s by their approximate posterior expectations using Bayes'
rule in (\ref{postprob}).
A second problem stems from the modeling of the putative variables as random
effects. The complete data
log-likelihood (\ref{eq:log.lik}) contains a large ($N\times N$) matrix of the form
$\mathbf{I}_N+\frac{\sigma^2}{\sigma_e^2}\mathbf{Z\bm{\Gamma}^2Z}'$, which has to be inverted
to compute the iterative maximum likelihood estimates.  
However, using the Woodbury identity \citep{Golub:1996}  $\log f_C(\mathbf{y},\bm{\Gamma})$ 
can be
expressed in terms of the $K\times K$ matrix
$\Sigma_K^* =
\mathbf{I}_K+\frac{\sigma^2}{\sigma_e^2}\bm{\Gamma}'\mathbf{Z}'\mathbf{Z}\bm{\Gamma}.$
This simplifies the computation because the $(k,l)th$ element of
$\bm{\Gamma}'\mathbf{Z}'\mathbf{Z}\bm{\Gamma}$ is given by
$\langle\mathbf{z}_k,\mathbf{z}_l\rangle\gamma_k\gamma_l$, where
$\langle,\rangle$ denotes the inner product of two vectors.  In contrast, the
elements of $\mathbf{Z\bm{\Gamma}^2Z}'$ involve all the $\gamma_ks.$
We set $\gamma_k^{(t)}=0$ if the posterior expectation of the $k^{th}$ latent
variable in the $t^{th}$ iteration is below a given threshold, and since only a
small number of the putative variables are truly associated with the response,
the matrix $\Sigma_K^*$ is very sparse and much easier to invert.
Thus, to obtain the inverse matrices of
$\bm{\Sigma}$ and $\mathbf{W}'\bm{\Sigma}^{-1}\mathbf{W}$ when
$N$ is large, we use the following form of $\bm{\Sigma}^{-1}$:
\begin{eqnarray}\label{SigmaK}
\bm{\Sigma}_{K}^{-1}\equiv
\bm{\Sigma}^{-1}=\frac{1}{\sigma_e^2}\mathbf{I}_N-
\frac{\sigma^2}{\sigma_e^4}\mathbf{Z}\bm{\Gamma}\left(\mathbf{I}_K+\frac{
\sigma^2}{\sigma_e^2}
\bm{\Gamma}\mathbf{Z}'\mathbf{Z}\bm{\Gamma}\right)^{-1}\bm{\Gamma}\mathbf{Z}'.
\end{eqnarray}
This implies we can rewrite the log-likelihood function:
\begin{eqnarray}\label{log.likelihood.K}
\ell &=&c_0\log(p_{0})+c_1\log(p_{1})+
c_2\log(p_{2})-\frac{N}{2}\log\left( 2\pi
\right)\nonumber\\
&&-\frac{N}{2}\log\left(\sigma_e^2\right)-\frac{1}{2}\log\left\vert
\mathbf{I}_{K}+\frac{\sigma^2}{\sigma_e^2}\bm{\Gamma}\mathbf{Z}'
\mathbf{Z}\bm{\Gamma} \right\vert \nonumber\\
&&-
\left( \mathbf{y}-\mathbf{X}\bm{\beta}-\mathbf{V} \bm{\mu }\right)'
\bm{\Sigma}_{K}^{-1}
\left( \mathbf{y}-\mathbf{X}\bm{\beta }-\mathbf{V}\bm{\mu }\right)\,,
\end{eqnarray}
where we have used the identity
$\left\vert I_{K}+AB^{T}\right\vert =\left\vert I_{N}+B^{T}A\right\vert$.

Suppose that there are $L$ variables for which $\gamma_k\neq 0$, and let
$\bm{\Gamma}_L$ be the corresponding $L\times L$ matrix ($\bm{\Gamma}_L$ is
obtained by eliminating all the 0 columns and rows in $\bm{\Gamma}$).  Let
$\mathbf{Z}_L$ be the sub-matrix obtained by eliminating the $K-L$ columns that
correspond to $\gamma_k= 0$, then we can rewrite (\ref{SigmaK}) as
\begin{eqnarray}\label{SigmaL}\bm{\Sigma}_{K}^{-1}\equiv
\bm{\Sigma}^{-1}=\frac{1}{\sigma_e^2}\mathbf{I}_N-
\frac{\sigma^2}{\sigma_e^4}\mathbf{Z}_L\bm{\Gamma}_L\left(\mathbf{I}_L+\frac{
\sigma^2}{\sigma_e^2}
\bm{\Gamma}_L\mathbf{Z}_L'\mathbf{Z}_L\bm{\Gamma}_L\right)^{-1}\bm{\Gamma}
_L\mathbf{Z}_L'\,.
\end{eqnarray}

Then as a result of applying Woodbury's identity and setting most $\gamma_k$ to
0, updating equations (\ref{mle.eq.beta}), (\ref{sigma_e}), (\ref{sigma}), and
 (\ref{log.likelihood.K}), with $\mathbf{V}=\mathbf{Z}_L\bm{\Gamma}_L$ and
$\mathbf{W}=\left[\mathbf{X},\mathbf{Z}_L\bm{\Gamma}_L\mathbf{1}_L\right]$, is computationally much simpler since it involves the
inversion of $L\times L$ matrices, where $L$ is much smaller than $N$ and $K$.

\section{Implementation Considerations}\label{sec:implementation}
\subsection{Model-related Issues}
\textbf{Correlated putative variables}:
One of the serious challenges in multiple linear regression is how to deal
with multicollinearity. When two or more of the predictors are highly correlated
the standard deviation of the parameter estimates may be severely inflated,
resulting in lack of power to identify potentially important variables.
Thus, multicollinearity can have a detrimental effect on variable selection methods.
A sequential method which does not account for possible correlations among
variables can lead to instability in the selection process, or to underfitting.
For example, if both $\mathbf{z}_1$ and $\mathbf{z}_2$ are associated with the response, but are also
highly correlated with each other, the selection procedure might add $\mathbf{z}_1$ first,
then $\mathbf{z}_2$, and then,
if due to the variance inflation the p-value of both variables becomes too large,
the procedure might exclude both from the model.
When $K>N$ the putative variables are necessarily correlated, and multicollinearity
simply cannot be avoided.

When a putative variable $\mathbf{z}_k$ that is not yet in the model after $t$ iterations is considered
its correlation with all
the variables currently in the model is computed.
If the maximum correlation is below a pre-defined threshold and the new variable yields
a significant improvement to the likelihood, then it is added to the model. Alternatively,
 the posterior probabilities of the
candidate variable can be adjusted as follows. Let
$C_k=\max_{j}\left\{cor(\mathbf{z}_k,\mathbf{z}_j)^2\right\}$ where $\left\{\mathbf{z}_j\right\}$
are the variables currently in the model.
Denote the unadjusted posterior probabilities by $P_s(k)$ for $s=-1,0,1$. The adjusted
posterior probabilities are $\tilde P_{-1}(k)=(1-C_k)P_{-1}(k)$,
$\tilde P_1(k)=(1-C_k)P_{1}(k)$, and $\tilde P_0(k)=1-\tilde P_{-1}(k)-\tilde P_{1}(k)$.
This shrinks the non-null posterior probabilities by a factor of $1-C_k$ which is close to
0 when $\mathbf{z}_k$ is correlated with any of the variables in the current model.

\justify\textbf{Compositional covariates}:
\cite{Lin13082014} discuss variable selection in regression with compositional
covariates. That is, the putative variables consist of proportions that sum to 1.
It is possible to modify the model in Section \ref{sec:vs:model} in order to account for
the sum constraint. However, it is equivalent to using the log-ratio transformation in
\cite{Aitchison:1982} in which the matrix $\mathbf{Z}$ is replaced with the $N\times K-1$
log-ratio matrix $\mathbf{Z}^K=[\log(z_{ij}/z_{iK})]$. Of course, any column in $\mathbf{Z}$
can be used as the reference component in the denominator, as long as it does not contain zeros.
This approach allows us to use our model directly, without any modifications.
We demonstrate it in the Case Studies section, using the data set from \cite{Lin13082014}.

\justify\textbf{Prior information on a subset of putative variables}:
In some situations we may wish to incorporate external
information about the putative variables. For example, when more
information is available about certain variables and how they might
be associated with the response we can account for that in the
prior probability distribution.  One way do it is to let $p_0,p_1,$ and $p_2$
depend on covariates and estimate them with a multinomial logistic regression
model. Alternatively, we can partition the covariates and assign each subset
a different multinomial prior.

\justify\textbf{Interactions between putative variables}:
Including interaction terms in this framework is straightforward.  To add an
interaction between $\mathbf{u}_k$ and $\mathbf{u}_m$ we simply augment
$\mathbf{Z}$ by adding a column which contains the element-wise product of the
$k^{th}$ and $m^{th}$ columns. Of course, including all pairwise interactions
of putative variables incurs a significant computational cost.

\subsection{Computational Issues}
\textbf{Memory requirements}:
When $K$ is very large it may not be possible to load the data matrix
$\mathbf{Z}$ to memory (let alone performing any computation and estimation).
Our model provides a way to avoid loading $\mathbf{Z}$ at once, since in each
step we only use a small subset of its columns.
Thus, we store $\mathbf{Z}$ on the hard disk, which has a much larger capacity
and just maintain an index of the columns so that the algorithm can easily and
quickly `fetch' only the necessary columns for each iteration of the algorithm.
Clearly, using the hard disk is slower than keeping the matrix in the computer's
memory, but using solid-state technology and choosing efficient storage and
indexing methods we can achieve excellent performance.

\justify\textbf{Large K}:
The most time-consuming part of the approximate EM algorithm is  the E-step
in which we compute $K$ posterior probabilities. Fortunately, this step can
be run in parallel since we fix the value of the model parameters that we obtained
in the previous M-step. Thus, using a cluster of computational nodes (e.g. cloud
computing services) we can divide the $K$
computations across $m$ nodes and we can expect to decrease the computational time
 by (approximately) a factor of $m$.

\justify\textbf{Higher-order terms}:
When higher-order term are included in the model we have
  found that rescaling all the variables so that their ranges are
contained in the interval $[-1,1]$, improves the stability of the estimation procedure.

\section{Simulations}\label{sec:sim}
We conducted a simulation study to verify that under the assumed model (\ref{model_1})
the algorithm yields accurate parameter estimates, and to compare the performance of our
algorithm in terms of power and accuracy with other methods.
Not surprisingly, as the sample size increases (even when $N$ is still much smaller than
$K$), the parameter estimates become more accurate, the power to detect the non-null
variables increases, and the Type-I error rate decreases. In this section, we describe the
results of a simulation study which compares the probabilities of Type I and Type II errors
and the coefficient of multiple determination, $R^2$,
with the well-known LASSO \citep{Tibshirani:1996}, as implemented in the \texttt{lars} package
\citep{Efron:2004, lars2013}.

The set-up in the simulations presented here is as follows.
We simulated 301 i.i.d. variables from a standard normal distribution. Variables 1-4 are all correlated,
with pairwise correlation coefficients $r_{ij}\approx 0.99$, and similarly for variables 5 and 6.
The response is determined by the formula $Y_i = Z_{3i} + Z_{6i} + Z_{7i} + Z_{8i} + \epsilon_i$ where$\epsilon_i\sim N(0,\sigma^2)$ independently for each $i$, and $\sigma^2=0.1$. We repeated the study with different
sample sizes ($N=40,60,80,100$) and here we describe the results for $N=40$.
To better demonstrate the differences between the methods in terms of $R^2$,
and to simulate missing variables in practical applications,  we drop $Z_8$, so
it is not available to the researcher in the fitting stage.
Note that dropping $Z_8$ has the same effect as increasing the error variance.
Thus, $K=300$.
The correlation between some of the explanatory variables means that any model that includes $Z_7$
and exactly one of $Z_1-Z_4$, and exactly one of $Z_5,Z_6$, will be a good fit for the data,
so as to avoid multicollinearity. So in that sense,
the `correct' number of variables in the model (or the number of true positives) is 3.
Note, however, that it is possible that by chance a small number of variables from $Z_9$ to $Z_{301}$
may actually contribute significantly to explaining variability in the response.

We used the greedy version of our algorithm and excluded a variable from the model if its final
posterior null probability, $p_{0}(k)$, was greater than 0.8. The results were not
sensitive to this threshold, since most null variables had posterior null probability close to 1.
When using the LASSO for variable selection we selected the tuning parameter, $\lambda$, for which
the 10-fold cross-validation mean squared error is minimized. Cross validation was repeated 30
times.

We repeated this simulation 100 times, and in each run we obtained the following: the $R^2$ of the
true model (including variables 3, 6, and 7), the $R^2$ from the model obtained using our method, and
the median $R^2$ from the 30 cross validation repetitions using the LASSO.
For each simulated data set we also counted the number of variables from the set 1-7,
and the number of variables from the set 9-301 which were selected by each method.

\begin{figure}[t!]
\begin{center}
\centerline{\includegraphics[trim=0mm 0mm 0mm 10mm,clip,width=6in]{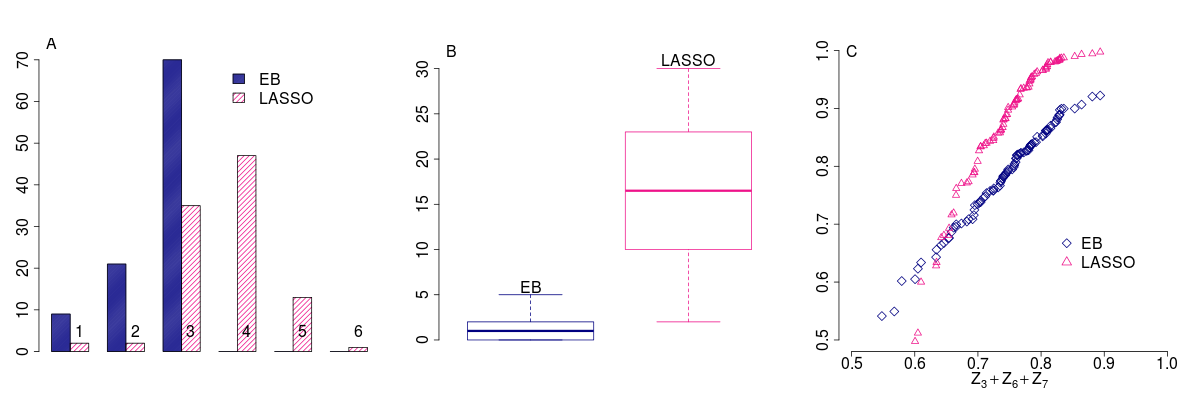}}
\caption{Simulation study -- empirical Bayes method vs. the LASSO. Left (A): the number of `true positives'
selected by each method (from the set $Z_1-Z_7$). Recall that only one of $Z_1-Z_4$ and only one of $Z_5-Z_6$
should be included in the model, in order to avoid multicollinearity.
Middle (B): The number of false positives obtained from each method.
Right (C): The $R^2$ values obtained by each method vs. the true $R^2$ when fitting the correct model, $Y=Z_3+Z_6+Z_7$.
\label{sim05}}
\end{center}
\end{figure}

Figure \ref{sim05} summarizes the differences between our empirical Bayes (EB) approach and the LASSO.
Panel A (left) shows the number of variables among $Z_1-Z_7$ that were selected in each run. Recall that
the high correlation between variables implies that there should be three variables from this subset
in the model, and any additional variable will introduce a multicollinearity problem.
The EB approach never had more than three variables from the set $Z_1- Z_7$ ($Z_7$, as well as
one from $Z_1-Z_4$, and one from $Z_5-Z_6$), and 70\% of the time
it had the correct number of variables. The LASSO, on the other hand, selected up to 6 of the 7 variables,
and over 60\% of the times the selected model had too many variables from the set $Z_1-Z_7$.

Panel B (middle) shows the number of variables from the set $Z_9-Z_{301}$ that were selected in each model.
The LASSO had a large number of `false positives' with a median of 18.5, whereas our EB approach had a
small number, with a median of 1. Note that in most cases the small number of false positives
that were included in models by our method were indeed significant when we fitted the linear model.
As mentioned earlier, in the simulation it can happen that some variables which were intended to be
`null' actually end up explaining a fairly large percent of the variability in the response, just by chance.

In the right panel (C) the $R^2$ values obtained from both methods are plotted versus the true $R^2$ values
(on the x-axis) which are obtained when we fit the correct model,
$Y_i = Z_{3i} + Z_{6i} + Z_{7i} + Z_{8i} +
\epsilon_i$.
The blue diamonds and the pink triangles represent the EB approach and the LASSO, respectively.
The LASSO inflates the $R^2$ by including so many false positive variables in the model.
In contrast, our method yields values which are approximately equal to the true values. Our $R^2$
values are typically slightly higher than the `true' $R^2$ because when our method selected variables
from the `null' set, these additional variables were actually significant when we fit the linear model.

Using the same simulation configuration, but with $N=80$ our empirical Bayes approach always selects exactly three variables from the non-null set.
In contrast, the LASSO selected three true positives 22\% of the time, and 78\% of the time
it selected 4-7 non-null variables (yielding a model with multicollinearity).

The number of false positives resulting from our model decreases with the sample size, and when $N=80$,
78\% of the runs having no false positives, 14\% have one false positive, 6\% have two false positives, 
and 2\% have 3.
The number of false positives resulting from the LASSO was very high and similar to the case of $N=40$ (as
depicted in Figure \ref{sim05}, B) with a median of 19.

\section{Case Studies}\label{sec:vs:cases}

\subsection{The Riboflavin Data}
In a recent paper demonstrating modern approaches to high-dimensional statistics,
\cite{buhlmann2014} used a data set in which the response
variable is the logarithm of riboflavin (vitamin B12) production rate,
and there are normalized expression levels of 4,088 genes which are used
as explanatory variables. The sample size in this data set is $N=71$.
In addition to the fact that the number of putative variable greatly exceeds
the number of observations, many of the putative variables are highly correlated.
Out of 8,353,828 pairs of genes, there are 70,349 with correlation coefficient
greater than 0.8 (in absolute value).

\cite{buhlmann2014} report that the Lasso with
subsamples of size $B=500$, and with $q=20$ variables that enter the regularization step
first, yields three significant and stable genes: LYSC\_at, YOAB\_at, and YXLD\_at.
The model with these three variables has an adjusted $R^2$ of 0.66, and AIC of 118.6.
Their multisample-split method yields one significant variable (YXLD\_at), and the
projection  estimator, used with Ridge-type score yields no significant variables
at the FWER-adjusted 5\% significance level.
The model with just YXLD\_at has an adjusted $R^2$ of 0.36 and AIC of 162.

\cite{LedererM15} also used this data set, to demonstrate their TREX model. Their final model
includes three genes: YXLE\_at, YOAB\_at, and YXLD\_at. The adjusted $R^2$ of their
model is 0.6, and the AIC is 130.68. Two of the genes are highly correlated (YXLD\_at and
YXLE\_at) and yield a variance inflation factor of 23.7 each, so when fitting the final
linear regression model neither appears to be significant.

\begin{figure}[t!]
\begin{center}
\centerline{\includegraphics[trim=0mm 0mm 0mm 10mm,clip,width=5in]{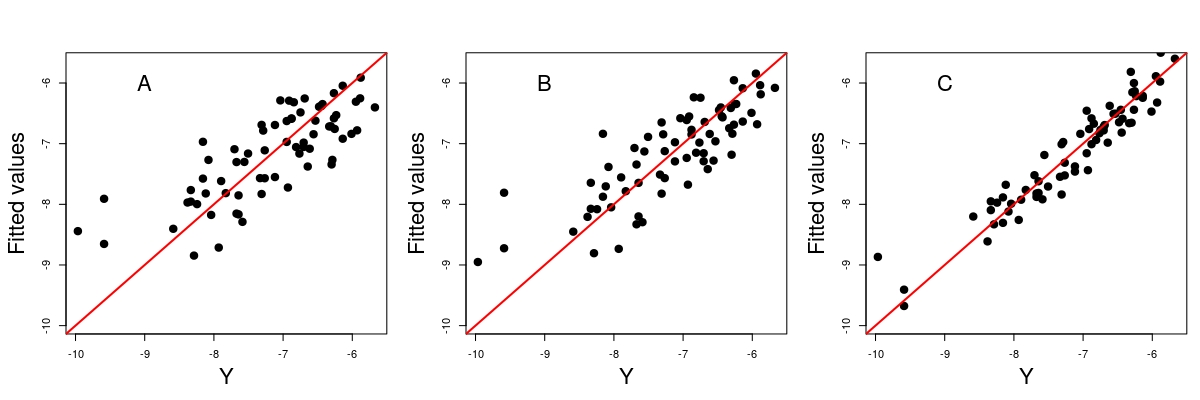}}
\caption{Riboflavin data -- fitted vs. observed values. A: \cite{LedererM15}, B: \cite{buhlmann2014},
C: our `weighted probability' (best fit).\label{riboflavin.aic}}
\end{center}
\end{figure}

We ran our algorithm using both the greedy and the weighted probability methods as
described in Section \ref{subsec:selection}.
The greedy method yielded five significant genes, with AIC=58.83  and an adjusted
$R^2$ of 0.86. We ran the algorithm using the weighted probability method 100 times.
The best model included seven genes, and had an AIC of 39.2 and an adjusted $R^2$
of 0.895.
The median AIC among the 100 fitted models was 84.38, and the maximum AIC was 114.1.
In summary, both methods yielded better fitting models than the ones obtained by
\cite{buhlmann2014} and \cite{LedererM15}. Table \ref{table:riboflavin:gof}
provides goodness-of-fit statistics for the different methods, including the
Aikake Information Criterion (AIC), the coefficient of multiple determination ($R^2$), and
the mean absolute error (MAE). Note that our best model explains 90.5\% of
the variability in the data.
Figure (\ref{riboflavin.aic}) shows the observed values (Y) vs. the fitted values from
three models: \cite{LedererM15}, \cite{buhlmann2014}, and our `weighted probability' (best fit).
In addition to having much smaller residuals than the two other methods,
our method provides much better prediction for low values of riboflavin. The other two
methods seem to  over-estimate the riboflavin levels when the true (normalized) values
are small (less than $-9$).

\begin{table}[ht]
\begin{center}
{\renewcommand{\arraystretch}{0.5}
\begin{tabular}{l|c|c|c}
 Method                            & AIC     & $R^2$ & MAE\\
 \hline
 \cite{LedererM15}                 & 130.681 & 0.616 & 0.462\\
 \cite{buhlmann2014}               & 118.625 & 0.676 & 0.411\\
 `greedy'                          & ~58.828 & 0.879 & 0.244\\
 `weighted probability' (best fit) & ~39.223 & 0.905 & 0.219\\
 \hline
\end{tabular}}
\caption{Goodness of fit statistics for the different methods - the riboflavin data. }
\label{table:riboflavin:gof}
\end{center}
\end{table}

Table \ref{table:riboflavin:bst} provides the parameter estimates for the best model
which was obtained using our weighted probability method. All seven variables
are significant and they all have low variance inflation factors,
indicating that the selected variables are not highly correlated.

\begin{table}[hb]
\centering
{\renewcommand{\arraystretch}{0.5}
\begin{tabular}{lrrrrr}
  \hline
 & Estimate & Std. Error & t value & Pr($>|t|$) & VIF \\
  \hline
(Intercept) & -7.60 & 0.06 & -135.32 & 0.00 & 1.00 \\
  ARGF\_at &  -0.86 & 0.09 &   -9.28 & 0.00 & 1.24 \\
  XHLB\_at &   0.69 & 0.16 &    4.31 & 0.00 & 2.90 \\
  XKDN\_at &  -0.51 & 0.15 &   -3.43 & 0.00 & 3.14 \\
  XKDS\_at &   0.43 & 0.19 &    2.21 & 0.03 & 3.77 \\
  YHDZ\_at &   0.46 & 0.09 &    5.21 & 0.00 & 1.41 \\
  YOAB\_at &  -1.13 & 0.10 &  -11.29 & 0.00 & 1.35 \\
  YXLE\_at &  -0.92 & 0.09 &  -10.39 & 0.00 & 1.41 \\
  \hline
\end{tabular}}
\caption{Parameter estimates for the best fitting model - the riboflavin data.}
\label{table:riboflavin:bst}
\end{table}

Clearly, our method found an excellent combination of a small number of putative variables
that explain over 90\% of the variability in the data, and none of these putative variables
is approximately a linear combination of the others.
However, one should be careful not to interpret it as the `right model' for two reasons.
First, when the number of putative variables is large there is no way to evaluate all possible
models. Second, if two genes are highly correlated and one of them is found to be significantly
associated with the outcome, then we expect the other gene to also be associated
with the outcome. Thus, we recommend to run the randomized version of our algorithm and
analyze the relationships between selected variables, as we demonstrate now.

Among the 191 variables that were selected by our method in at least one iteration
of the weighted probability method there are 64 highly correlated pairs (absolute value of the
correlation coefficient greater than 0.8). Now consider the seven variables in the best fitting model.
Table \ref{table:riboflavin:bst.cor} shows the number of models in which each of these
variables was included. Three of the variables actually represent a set of highly correlated genes.
For example, ARGF\_at was selected 8 times, but it is highly correlated with three other genes
(ARGD\_at, ARGH\_at, and ARGJ\_at) which were selected a total of 4 times. So this set of related genes
was selected in 12 of the models.
The cluster of genes which are correlated with XHLB\_at consists of 7 genes and was
represented in 95 of the models.
Similarly, YXLE\_at represents a cluster of 6 genes, of which one was included in 97 of the models.
The large number of models which included a variable from the XHLB\_at and YXLE\_at clusters
suggests that the final model should include a representative from these clusters.
Note that our model never selected more than one gene from the same cluster.
The other four selected genes were not correlated with any other selected genes.
Two of these genes (XKDS\_at and YOAB\_at) appear in a relatively large number of models
(23 and 24, respectively).

\begin{table}
\centering
{\renewcommand{\arraystretch}{0.5}
\begin{tabular}{lc|ll|c}
  Variable & \# models & Correlated variables & \# models & Total models\\
  \hline
  ARGF\_at &  8       & ARGH\_at, ARGJ\_at, ARGD\_at & 2, 1, 1 & 12 \\ [1ex]
  \hline
  XHLB\_at &  22      & XKDF\_at, XKDH\_at, XKDI\_at, &   13, 2, 19, &  95\\
   &       & XKDK\_at, XHLA\_at, XLYA\_at &    9, 17, 13 & \\
  \hline
  XKDN\_at &  1       &  &    & 1 \\
  \hline
  XKDS\_at &  23      &  &    & 23 \\
  \hline
  YHDZ\_at &  3       &  &    & 3 \\
  \hline
  YOAB\_at &  24      &  &    & 24 \\
  \hline
  YXLE\_at &  18      & YXLC\_at, YXLD\_at, YXLF\_at,   & 6, 18, 23,  & 97 \\
   &       & YXLG\_at, YXLJ\_at & 14, 18 & \\
  \hline
\end{tabular}}
\caption{The riboflavin data - the number of models in which each selected variable was included.
ARGF\_at, XHLB\_at, YXLE\_at are correlated with other selected genes. The last column shows the total
number of models (out of 100) in which each cluster of genes was represented. }
\label{table:riboflavin:bst.cor}
\end{table}

\subsection{The BMI Data}
\cite{Lin13082014} demonstrated an application of a LASSO-based variable selection
method for regression models with compositional covariates. The analysis aims
to identify a subset of 87 bacteria genera in the gut whose subcomposition
is associated with body-mass index (BMI). The data in this case is compositional,
which using our previous notation means that $\sum_{j=1}^{87} z_{ij}=1$ for each $i$.
To apply our method directly, without changing the model to account for the sum
constraint, we perform the log ratio transformation and replace the matrix $\mathbf{Z}$
with $\mathbf{Z}^K=[\log(z_{ij}/z_{iK})]$. The data contains many zero counts, so
\cite{Lin13082014} replace them with 0.5 before converting the data to be in
compositional form. 
We analyzed the data with all 87 bacteria
and obtained a model that includes all four variables in \cite{Lin13082014} and two additional genera (Dorea and Oscillibacter). \cite{wu:2011} report that the \emph{Oscillibacter} genus was
negatively correlated with BMI,  which agrees with our overall
model. However, when we refit the six genera model we find that
\emph{Oscillibacter} is no longer significant ($p= 0.657$).

In this section we summarize the results of our analysis using a subset of
45 bacteria which had non-zero counts in at least 10\% of the samples ($N=96$).  The
omitted genera have minimal contribution to the overall distribution of
the proportions. The \cite{Lin13082014} analysis finds four genera (\emph{Acidaminococcus, Alistipes, Allisonella, Clostridium}).
The best model obtained by our algorithm includes 8 genera and it yields
an increase of 7.5\% in the
amount of variability explained by the model when compared to the model obtained by
\cite{Lin13082014}. However, when fitting the
regression model with the selected genera we observe that two of them (\emph{Dorea} and
\emph{Oscillibacter}) are not significant.
We fit a reduced model with 6 genera (\emph{Acidaminococcus, Alistipes, Allisonella, Catenibacterium, Clostridium, Megamonas}) and obtain the lowest AIC (564.5) and the largest adjusted $R^2$ (0.345).
Some goodness of fit statistics for the different models can
be found in Table \ref{table:bmi:gof}, and the parameter estimates for the final model
are provided in Table \ref{table:bmi:parest}.

\begin{table}[ht]
\begin{center}
{\renewcommand{\arraystretch}{0.5}
\begin{tabular}{l|c|c|c|c}
 Method                                 & Variables & AIC     & $R^2$ & MAE  \\
 \hline
 \cite{Lin13082014}                     & 4         & 569.7 & 0.324 & 3.290\\
 `weighted probability'                 & 8         & 566.5 & 0.399 & 3.123\\
 `weighted probability' (reduced model) & 6         & 564.5 & 0.386 & 3.158\\
 \hline
\end{tabular}}
\caption{Goodness of fit statistics - the BMI data. }
\label{table:bmi:gof}
\end{center}
\end{table}

\begin{table}[hb]
\centering
{\renewcommand{\arraystretch}{0.5}
\begin{tabular}{lrrrr}
 & Estimate & Std. Error & t value & Pr($>|t|$) \\
  \hline
(Intercept) & 28.41 & 1.49 & 19.01 & 0.00 \\
  \emph{Alistipes} & -0.67 & 0.25 & -2.73 & 0.01 \\
  \emph{Clostridium} & -1.04 & 0.31 & -3.33 & 0.00 \\
  \emph{Acidaminococcus} & 0.92 & 0.25 & 3.59 & 0.00 \\
  \emph{Allisonella} & 1.39 & 0.56 & 2.49 & 0.01 \\
 \emph{ Megamonas} & -0.86 & 0.33 & -2.61 & 0.01 \\
  \emph{Catenibacterium} & 0.74 & 0.36 & 2.06 & 0.04 \\
   \hline
\end{tabular}}
\caption{Parameter estimates in the BMI data. }
\label{table:bmi:parest}
\end{table}

Our results are consistent with the previous findings in the microbiome literature.  The relative proportion of phylum \emph{Bacteroidetes} (containing the \emph{Alistipes} genus) to phylum \emph{Firmicutes} (all others genera found) is lower in obese mice and humans than in lean subjects \citep{ley:2006}. Furthermore the family \emph{Veillonellaceae} (containing the \emph{Acidaminococcus} and \emph{Allisonella} genera) are positively correlated to BMI \citep{wu:2011}.  The two additional genera found beyond \cite{Lin13082014}, \emph{Catenibacterium} and \emph{Megamonas}, have been found to be associated with BMI.   \cite{chiu2014}  identifies \emph{Megamonas} as a genus that differentiates between low and high BMI in a Taiwanese population and  \cite{turnbaugh2009} found that the \emph{Catenibacterium} genus increased diet-induced dysbiosis for high-fat and high-sugar diets in humanized gnotobiotic mice.

\section{Extensions}\label{sec:varsel:ext}
We briefly discuss a number of useful extensions to our model.

\textbf{Mixed Models}:
In model (\ref{model:varsel}) we assumed that the mean of the
response is a linear combination of two groups of variables, where
$\mathbf{X}\bm{\beta}$  represents the `locked' variables, and
$\mathbf{Z}\bm{\Gamma}\mathbf{u}$  represents the putative variables.
We assumed that the `locked' variables are treated as fixed effects in the model.
There are situations in which we may want to `lock' additional \textit{random}
effects in the model.
For example, in biological applications (e.g. QTL analysis) one may want to
include breed, or kinship information as a random effect.
This can be done using the same method that we used to estimate the
variance parameter, since the update equations for the EM algorithm extend to
any number of variance components (see the general formulation of the estimation
in Section 8.3.b in \citealt{McCulloch:1992}).  For example, with one additional
random effect only one parameter is added to the model, which can be modified as follows:
\begin{eqnarray*}
\mathbf{y}
&=&\mathbf{X}\bm{\beta}+\mathbf{Z}_0\mathbf{v}+\mathbf{Z}\bm{\Gamma}\mathbf{u}
+\bm{\varepsilon}\\
\bm{\varepsilon } &\sim &N\left( \mathbf{0}_{N},\sigma _{e}^{2}%
\mathbf{I}_{N}\right) \\
\mathbf{Z}_0\mathbf{v} &\sim &N\left( \mathbf{0}_{N},\sigma_0
^{2}\mathbf{I}\right)\\
\mathbf{Z}\bm{\Gamma}\mathbf{u} &\sim &N\left(
\mathbf{Z}\bm{\Gamma}\bm{\mu},\sigma ^{2}\mathbf{Z}\bm{\Gamma}^2%
\mathbf{Z}'\right)\,.
\end{eqnarray*}

\textbf{Interactions}:
In many applications it may be useful to include interactions between
fixed effects and all putative variables. For example, if the response is
the Forced Expiratory Volume (FEV, which measures how much air is exhaled during
a forced breath), it may be the case that the effects of interest are the
interactions between gene expression levels (the putative variables) and
smoking status (the fixed effects). It is possible that the genetic variables
have a different effect on the response for heavy smokers than for non-smokers.

\textbf{Generalized Linear Models}:
Our model can be extended to the generalized linear model framework.
This will allow us to deal with binomial and Poisson
responses, as well as censored survival times using the artificial Poisson model
as described by \cite{Whitehead:1980}.
Suppose the response mean vector,
$\bm{\mu}$, is related to the linear predictor, $\bm\eta=\mathbf{X}\bm\beta+\mathbf{Z}\mathbf{u}$,
via a link function, $g(\bm\mu)=\bm\eta$, and conditional on the latent
indicator vector, $\bm\gamma$, the responses are independent outcomes
from an exponential dispersion family; i.e.
\begin{eqnarray}
 \label{eq:glmm.model}
  f(y_i|\mathbf{u},\bm\gamma)=h(y_i)\exp\left\{\frac{w_i}{\phi}\left[ \theta_iy_i-b(\theta_i)\right]+c(y_i;\phi/w_i)\right\}
\end{eqnarray}
where the identity $\mu_i=b'(\theta_i)$ relates the canonical
parameter for the $i$th response to its mean, $\phi$ is a (possibly
known) dispersion parameter, and $w_i$ are known weights.
In this setting implementation of the EM algorithm is
further complicated by the fact that the complete data likelihood for
$(\mathbf{y},\bm\gamma)$ is not analytically tractable, with the exception
being the normal response case, where it is given in
(\ref{log.likelihood}).

An approximation of the intractable likelihood can be developed by using
penalized quasi-likelihood (PQL) approaches proposed by
\citet{scha:1991}, \citet{bres:clay:1993} and
\citet{wolf:ocon:1993}. The basic idea of the PQL approach is to
replace the generalized linear mixed model by an approximating
linear mixed model, $\mathbf{z}=\mathbf{X}\bm\beta+\mathbf{Zu}+\mathbf{e}$,
in which $\mathbf{e}=(\mathbf{y}-\bm\mu)(\partial\bm\eta/\partial\bm\mu)\sim
N(0,\mathbf{W}^{-1})$, $\mathbf{u}\sim N(\psi\mathbf{1}_M,\sigma^2_u\mathbf{I}_K)$,
and $\mathbf{W}=diag[w_i(\partial\mu_i/\partial\eta_i)^2/\phi V(\mu_i)]$, where $V(\cdot)$
denotes the variance function for the GLM. For fixed $\bm\gamma$,
estimation of the model parameters proceeds in a similar manner to
that for a normal theory linear mixed model, except that the {\it
  working response} vector, $\mathbf{z}$, must be iteratively updated. There
are also a several different algorithms for estimating the variance
components $(\phi,\sigma^2_u)$ \citep[see][]{pawi:2001}. The key
identity (\ref{SigmaK}) is replaced by
\begin{eqnarray*}
  \left( \mathbf{W}^{-1}+\sigma^2_u\mathbf{ZZ}' \right)^{-1}
  = \mathbf{W} + \mathbf{Z}\left(\frac{1}{\sigma^2_u}\mathbf{W} +
\mathbf{Z}'\mathbf{W}\mathbf{Z}\right)^{-1}\mathbf{Z}'\,,
\end{eqnarray*}
with the $(k,l)$th component of $\mathbf{Z}'\mathbf{W}\mathbf{Z}$ being
$(\mathbf{z'}_k\mathbf{W}\mathbf{z}_l)\gamma_k\gamma_l$, so that threshold based on the
posterior expected $\gamma$'s again results in a much lower dimensional
matrix inversion at the M-step.

Developing methods to implement these extensions will be considered in future work.

\section{Conclusion}\label{sec:varsel:conc}
In 1996 Brad Efron stated that variable selection in regression is the most important
problem in statistics \citep{Hesterberg:2010}. Since then many papers have been written
on the topic, as this continues to be a challenging problem in the age of high-throughput
sequencing in genomics, and as other types of `omics' data become available and more
affordable.

We developed a model-based, empirical Bayes approach to variable selection.
We propose a mixture model in which the putative variables are modeled as random effects.
This leads to shrinkage estimation and increases the power to select the correct variables,
while maintaining a low rate of false positive selections.
The mixture model involves a very small number of parameters. Importantly, that number
remains constant regardless of the number
of putative variables. This parsimony contributes to the scalability of the algorithm
and it can handle a large number of predictors. In contrast, methods based on
MCMC sampling often require the user to focus on a relatively small subset of the
predictors. Using the EM algorithm and an efficient dimension reduction
method our algorithm converges significantly faster than simulation-based methods.
Furthermore, a simple modification to the algorithm prevents multicollinearity problems in
the fitted regression model.

\bibliographystyle{ims}

\bibliography{refs}

\end{document}